\documentclass[pre,aps,twocolumn, floatfix, superscriptaddress,showpacs]{revtex4}
\usepackage{amsmath,bm,graphicx}
\usepackage{amssymb}
\usepackage{amsfonts}
\usepackage{times}
\usepackage{graphicx,color}
\definecolor{r}{rgb}{1,0,0}
\definecolor{b}{rgb}{0,0,1}

\begin{document}
\title{Fragility and hysteretic creep in frictional granular jamming}

\author{M. M. Bandi}
\affiliation{MPA-10, Los Alamos National Laboratory, Los Alamos, NM 87545, USA}
\affiliation{T-CNLS, Los Alamos, National Laboratory, Los Alamos, NM 87545, USA}
\affiliation{Current Affiliation: Collective Interactions Unit, OIST Graduate University, 1919-1Tancha, Onna-son, Kunigami-gun, Okinawa, Japan 904-0495}
\email[Corresponding Author: ]{bandi@oist.jp}

\author{M. K. Rivera}
\affiliation{D-4, Los Alamos National Laboratory, Los Alamos, NM 87545, USA}

\author{F. Krzakala$^2$}
\affiliation{CNRS and ESPCI ParisTech, 10 rue Vauquelin, UMR 7083 Gulliver, Paris 75000 France}

\author{R. E. Ecke$^2$}

\date{\today}
\begin{abstract}
The granular jamming transition is experimentally investigated in a two-dimensional system of frictional, bi-dispersed disks subject to quasi-static, uniaxial compression without vibrational disturbances (zero granular temperature). Currently accepted results show the jamming transition occurs at a critical packing fraction $\phi_c$ defined as the packing fraction $\phi$ at which the pressure rises above zero with power-law scaling and the system hits the isostatic point where all displacements come to a halt. In contrast, we observe the first compression cycle exhibits {\it fragility} - metastable configuration with simultaneous jammed and un-jammed clusters - over a small interval in packing fraction ($\phi_1 < \phi < \phi_2$). The fragile state separates the two conditions that define $\phi_c$ with an exponential rise in pressure starting at $\phi_1$ and an exponential fall in disk displacements ending at $\phi_2$. The results are explained through a percolation mechanism of stressed contacts where cluster growth exhibits strong spatial correlation with disk displacements. The source of fragility is traced to the experimental incompatibility between the requirements for zero friction and zero granular temperature, with the $\phi_c$ recovered for non-zero granular temperature. Measurements with several disk materials of varying elastic moduli $E$ and friction coefficients $\mu$, show friction directly controls the start of the fragile state, but indirectly controls the exponential slope. Additionally, we experimentally confirm recent predictions relating the dependence of $\phi_c$ on $\mu$. Under repetitive loading (compression), the system exhibits hysteresis in pressure, and the onset $\phi_c$ increases slowly with repetition number. This friction induced hysteretic creep is interpreted as the granular pack's evolution from a metastable to an eventual structurally stable configuration. It is shown to depend upon the quasi-static step size $\Delta \phi$ which provides the only perturbative mechanism in the experimental protocol, and the friction coefficient $\mu$ which acts to stabilize the pack.
\end{abstract}
\maketitle

\section{Introduction}
The ``Jamming'' framework \cite{LN1998} has been proposed as an overarching, unifying description governing the behavior of a wide variety of disordered materials including glasses, colloids, foams, and granular media via the jamming phase diagram with three axes representing temperature $T$, shear stress $\Sigma$, and inverse packing fraction $1/\phi$. Of particular interest was a point $J$ at zero temperature and zero shear stress along the inverse packing fraction axis ($1/\phi$) that is predicted to have special properties. This point $J$ is the critical packing fraction $\phi_c$ at which a granular pack undergoes a {\em sharp} transition from an athermal, particulate gas to a stiff, disordered solid. Numerics \cite{OSLN2003}, mean field theories \cite{HC2005}, and experiments \cite{MSLB2007} alike show many interesting properties arise at this transition. Being amenable to theoretical treatment, an overwhelming majority of the studies have concentrated on idealized systems composed of {\em frictionless} particles. From the perspective of real granular materials, however, the effects of inter-particle friction on the jamming scenario have not been considered \cite{Radjai1996, SEGHL2002, Somfai2007, Shundyak2007} as widely as the frictionless case \cite{SWM2008}. Nevertheless, friction plays a significant role in such materials; not only is it technologically relevant (e.g., in compression and sintering of ceramics \cite{Cumberland1987}), but friction also radically alters the mechanical behavior of materials (e.g., in sedimentary rocks \cite{Tutuncu1998}). Understanding the role of friction in granular jamming therefore becomes relevant and important.

In this article, we present an experimental study of a loose granular pack comprising a two-dimensional, bi-disperse system of disks subjected to quasi-static, uniaxial compression (loading) and decompression (unloading). The following experimental results and their interpretations are presented:

1) By employing disks with different static friction coefficients, we verify numerical predictions \cite{SEGHL2002, Silbert2010} for the effect of friction on the onset of jamming.

2) We show the pressure scaling is remarkably different between the first and subsequent loading cycles. In the first cycle as the system's boundaries are moved in to achieve an increasingly tighter packing, a fragile state is observed where the pressure exhibits an exponential rise $P \propto e^{\phi/\chi_P}$ over a range of packing fractions ($\phi_1 < \phi < \phi_2$), followed by a deviation from exponential scaling for $\phi > \phi_2$. This exponential rise in pressure is simultaneously reflected in an exponential decrease in particle displacements over the same range of packing fractions, implying the simultaneous existence of jammed and unjammed clusters in the evolving granular pack. This fragile state is characterized by developing contact stresses being spatially correlated with disk displacements. It is shown to arise as a consequence of non-zero friction which is known to introduce protocol dependence on experimental measurements. These results are interpreted as a percolation of stressed contacts which exponentially decrease the fractional area enclosed within stress chains (defined by a threshold stress) over the range of packing fractions ($\phi_1 < \phi < \phi_2$). The fragile state and its associated stress percolation mechanism fall within a broader set of recent results that show existence of two distinct regimes and point to some form of percolation mechanism in the approach to jamming including contact percolation \cite{SOS2012}, contact dynamics in granular glasses \cite{CBD2012}, and shear induced jamming that causes force network percolation \cite{BZCB2011}.

We note the term ``fragile'' here denotes a mechanically metastable configuration - it is easily destroyed under the slightest external perturbation such as non-zero granular temperature, as we show later in the article. This state should not be confused with fragility in glass formers as defined by Angell \cite{Angell1988} to distinguish between strong and fragile glasses which exhibit Arrhenius and Vogel-Fulcher behavior respectively \cite{Vilgis1993}. Although related to glasses within the broader context of disordered systems, this work does not concern itself with glass phenomenology in particular.

3) Under repetitive loading and unloading, we observe the critical packing fraction $\phi_c$ at which the granular pack jams progressively increases to higher values, thereby exhibiting creep. At the same time the pressure curves for the loading-unloading cycles exhibit hysteretic responses. Relevant in the geophysical context \cite{CKT2009, MeyersChawla}, this hysteretic creep is experimentally shown to arise from inter-particulate contact friction.

In section II, we present a brief review of frictionless granular jamming, and how friction changes the predictions expected to hold under idealized conditions. In Section III, we explain our experimental setup and the experimental protocol we follow in conducting our measurements. The main results of this study are presented in Section IV, followed by discussion and interpretation of these results in Section V, with a brief summary of results in Section VI.

\section{Background: Granular Jamming}
\subsection{Ideal Jamming}
The primary motivation for the jamming proposal \cite{LN1998} was to provide a common framework to describe the non-equilibrium behavior of a wide variety of disordered materials. O'Hern et al. \cite{OSLN2003} conducted extensive numerical studies with frictionless particles interacting via soft, finite-range, repulsive potentials at zero temperature and zero applied shear stress - henceforth we refer to this set of specifications as {\em Ideal Granular Jamming}. They reported many interesting properties of the ideal jamming transition around $\phi_c$, that have since been verified by mean field theories \cite{HC2005} and experiments \cite{MSLB2007}. For a finite number of particles $N$, $\phi_c$ is a configuration dependent random variable, the full-width at half-height of whose distribution was empirically determined \cite{OSLN2003} to follow the formula $w = w_0N^{-\Omega}$ (where $w_0 = 0.16 \pm 0.04$ and $\Omega = 0.55 \pm 0.03$). O'Hern et al. found $\phi_c$ is sharply defined in the limit of infinite system size ($N \rightarrow \infty$), where $\phi_c$ coincides with Random Close Packed density ($\phi_{RCP}$ = 0.64 in 3D and 0.84 in 2D), a concept first introduced by J. D. Bernal \cite{Bernal1960, BM1960, Scott1960} to understand liquids which are structurally disordered by construction. We note, however that divergence of opinion exists within the community both with respect to the definition of random close packing \cite{TTD2000} itself, as well as its coincidence with $\phi_c$ in granular jamming \cite{CBS2010, FK2007}. This divergent opinion notwithstanding, the existence of a $\phi_c$ where the jamming transition occurs is not in question.
 
The behavior of two quantities is of particular interest for the ideal jamming transition. The pressure $P$ is zero below $\phi_c$ and rises continuously above $\phi_c$ as a power law ($P \propto (\phi-\phi_c)^{\psi}$ for $\phi > \phi_c$), whereas the coordination number $Z = 0$ below $\phi_c$ and undergoes a discontinuous jump to a critical value $Z = Z_c$ at $\phi = \phi_c$, followed by a power law increase above $\phi_c$ ($(Z - Z_c) \propto (\phi - \phi_c)^{\beta}$). The critical coordination number $Z_c = 2D$ ($D$ being the system's dimensionality) for frictionless particles, since $\phi_c$ is the system's isostatic point at which the total number of degrees of freedom equal the total number of constraints providing force balance.

The definition of a contact plays a central role here. A granular contact is said to exist when two particles come in physical contact and propagate a stress; albeit necessary, a stress-free physical contact is deemed insufficient. This requirement gives rise to a crucial, and perhaps little appreciated interpretation of the jamming transition. The discontinuous jump in $Z$ from 0 to $Z_c$ at $\phi_c$ implies that there are no contacts in the system up to $\phi_c$ - it may lose floppy modes and become rigid, but it is not stressed and therefore has zero granular contacts. At $\phi_c$, a pressure rise above zero simultaneously activates contacts system wide owing to stress propagation, and $Z$ discontinuously jumps from 0 to $Z_c$. Because of the additional stress requirement in contact definition, granular jamming ideas make no allowance for a system to lie in an intermediate regime where part of the system is jammed and the rest is not - the only two states permitted are total system-wide jamming or lack thereof.

\subsection{Frictional Jamming}
A few studies \cite{SEGHL2002, Shundyak2007, Somfai2007, Radjai1996, Silbert2010} have explored the role of friction and how it changes the ideal jamming predictions. Although a general framework for frictional packs has yet to emerge, studies indicate deviations from ideal jamming predictions. First, in addition to the normal component of force ($F_N$), friction gives rise to a tangential component ($F_T$). Second, the preparation method and history of the pack become important for frictional packs \cite{IOS2011}. Treating the tangential force component as a new independent degree of freedom (at least for low friction coefficients) sets the critical coordination number at $\phi_c$ to lie in the range $D+1 \le Z_c \le 2D$, i.e., hypostatic configurations are permissible. The condition $Z_c < 2D$ implies that frictional packs can jam at $\phi_c < \phi_{RCP}$, with $\phi_c$ progressively decreasing with increasing $\mu$ before asymptoting to a constant value that has been termed Random Loose Packing density $\phi_{RLP}$ \cite{Silbert2010}. $\phi_{RLP}$ is an empirically determined value from numerical simulations; its theoretical underpinnings are not well understood \cite{SWM2008}. The same arguments made against the definition of $\phi_{RCP}$ \cite{TTD2000} apply to $\phi_{RLP}$.

\section{Experiment}

\begin{figure}
\begin{center}$
\begin{array}{cc}
\includegraphics[width = 3.4 in]{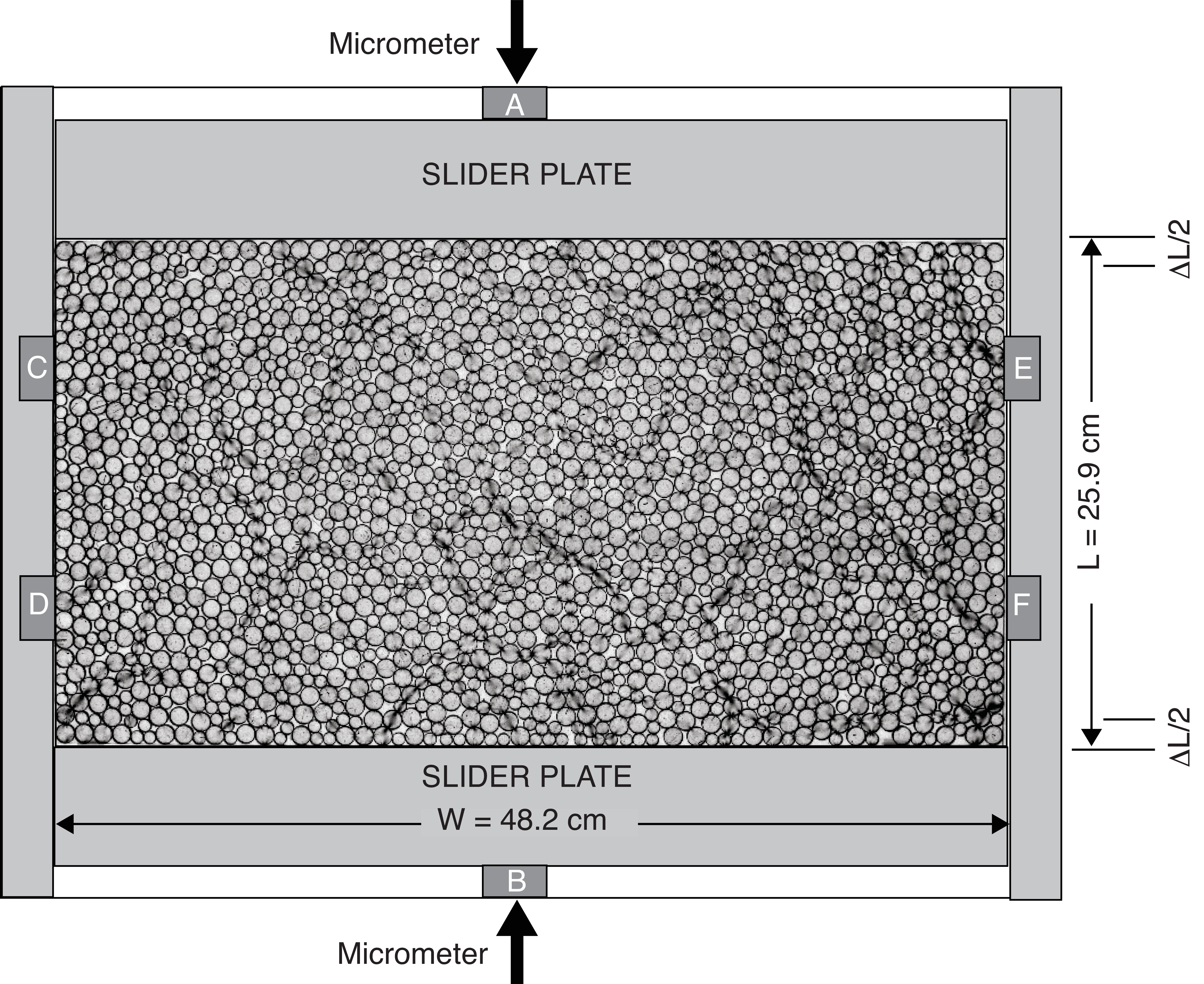}\\
\includegraphics[width = 2.5 in]{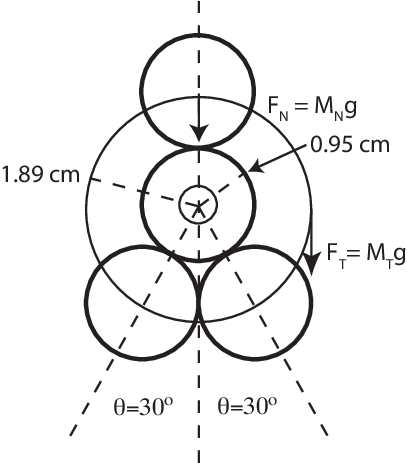}
\end{array}$
\end{center}
\caption{Top panel: Schematic of the experimental setup. The system consists of 950 large and 950 small disks (ratio of radii 1:1.5). Two movable boundaries at opposite ends control the system's packing fraction $\phi$, and are used to provide uniaxial, quasi-static compression. Force sensors labeled A through F in the schematic measure the boundary pressure. The image is contrast enhanced data for $\phi = 0.8113$. Bottom panel: Schematic of the static friction coefficient measurement apparatus. Four disks of diameter $d_L = 0.9525$ cm are placed in contact with each as shown in the schematic. The three outer disks are held fixed and have no translational or rotational degrees of freedom. The middle disk can be rotated by means of an external lever. A normal force $F_N = M_Ng$ is applied by a suspended weight of mass $M_N$ on the top disk. This force is transferred to the disk sandwiched in the middle which in turn transfers it to the two bottom disks at an angle of $30^{\circ}$. A tangential force $F_T = M_Tg$ applied on the middle disk allows it to slip at a critical force which allows determination of the static friction coefficient $\mu$ for the disks. }
\label{fig:fig1}
\end{figure}

\subsection{Description of the Experimental setup}
Figure \ref{fig:fig1} (top panel) shows a schematic of the experimental setup. The system consists of a bi-disperse mixture of 950 large (diameter $d_L = 0.9525 \pm 0.0025$ cm) and 950 small (diameter $d_S = 0.635 \pm 0.0025$ cm) disks of thickness 0.508 cm. The disks are placed in a chamber, with dimensions $L=  25.9$ cm and $W = 48.2$ cm ($L$ is the compression direction), consisting of two glass plates held 0.635 cm apart by means of an acrylic frame that runs along the system's perimeter. Two movable boundaries are placed on the acrylic frame with aluminum plates that can slide back-and-forth within the chamber from opposite ends. The transverse boundaries are held fixed. The positions of the movable boundaries are controlled by two micrometers with a precision of 0.001 cm. Taking variations in radii into account for the given system of 1900 disks, this translates to a precision $\Delta \phi = 1 \times 10^{-5}$ in the packing fraction, and hence serves as the lower bound on the quasi-static step-size ($\Delta \phi$). All measurements reported in this article, however, were made at a quasi-static step-size of $\Delta \phi = 1 \times 10^{-4}, 3.5 \times 10^{-4}$, or $7 \times 10^{-4}$. The packing fraction ($\phi$) is defined as the ratio of the area occupied by the disks to the total chamber area. The packing fraction is therefore controlled by changing the chamber area in this experiment. A set of six sensors (labeled A through F in fig.\ \ref{fig:fig1} top panel) placed along the boundaries measure the global two-dimensional pressure (N/m). The noise floor of these sensors is of the order of 0.1 N/m. Along the compression direction, the sensors feel the friction of the movable boundaries with the glass bottom, which was measured to be about 11 N/m, accordingly this serves as the true zero of pressure measured along the compression axis. Visual measurements using a Nikon D-90 camera (12.3 megapixel resolution) yield positions and displacements of individual disks.

In order to ascertain the magnitude of systematic error in measured global pressure introduced due to friction between the movable boundary and the bottom glass plate, we modified the arrangement of pressure sensors. In particular, rather than measure the pressure at a position between the compression micrometer and the movable boundary, two sensors (as opposed to one) were placed near a very small and light frame located near the disks.  The presence of a top glass confining plate was also considered.  The number and variety of runs with this modified apparatus were small compared to those taken with the primary apparatus whose results we discuss here. The general characteristics of the results are the same for both experiments.

Measurements were performed with disks machined from different materials spanning a range of experimentally-determined static friction coefficients. The majority of the measurements were conducted with polymer disks that exhibit stress birefringence or photoelastic response (PE, static friction coefficient $\mu = 0.19$, elastic modulus $E = 2.5$ GPa). Measurements were also performed with photoelastic disks lubricated with graphite dust (PEG, $\mu = 0.14$, $E = 2.5$ GPa), photoelastic disks soaked in Ethanol for 24 hours which changed the modulus (PEA, $\mu = 0.19$, $E = 0.004$ GPa), Lexan polycarbonate disks intentionally machined rough to obtain a high friction coefficient (LEX, $\mu = 0.22$, $E = 2$ GPa), and Teflon disks with an intrinsically low friction coefficient (TEF, $\mu = 0.06$, $E = 0.5$ GPa). Although photoelastic disks were employed to discern the spatial stress distribution, the photoelastic threshold of the material was found to be high owing to high stiffness of the photoelastic material used. As a result, photoelastic stress signals were not visible well into the rise in the system's global pressure measured with boundary sensors. Owing to this design shortcoming, we are unfortunately unable to provide measurements of the coordination number $Z$ in this article.

\subsection{Measurement of static friction coefficient $\mu$}
A schematic of the apparatus used to measure the static friction coefficient is shown in fig.\ \ref{fig:fig1} (bottom panel). Four disks (of the same material) of diameter $d_L = 0.9525$ cm are arranged as shown in fig.\ \ref{fig:fig1} (bottom panel). The upper disk and the two bottom disks are held fixed, and not allowed to rotate. A mass $M$ is suspended from the upper disk such that a force $F_a = Mg$ is applied vertically down by the upper disk onto the central disk. A tangential force $F_m = mg$ is applied on the central disk (free to rotate) via a pulley system where a mass $m$ is attached to the pulley with diameter $d_p = 1.8872$ cm. In practice, the weights are placed in a receptacle of mass 29 g, hence $F_m = (m+29) g$. Owing to the torque ratio, the actual tangential force applied on the central disk is $F_T = F_md_p/d$.

The normal force on the central disk arises from the three (one top and two bottom) disks in contact with it. The total normal force $F_a$ exerted by the top and central disks is balanced by the two bottom disks. Hence $F_a = 2F_N\cos\theta$, or $F_N = F_a(1+1/\cos\theta)$. The ratio of the total normal force applied to the total tangential force required for a slip to occur in the central disk provides a measure of the static friction coefficient $\mu$. Plugging in values for $d_P/d_L = 1.98$ and $(1+1/cos(\theta)) = 2.15$ with $\theta = 30\,^{\circ}$, we get:

\begin{equation}
\label{friction}
\mu = \frac{F_T}{F_N} = \frac{F_md_P/d}{F_a(1+1/cos(\theta))} = \frac{1.98F_m}{2.15F_a} = 0.92 \frac{F_m}{F_a}
\end{equation}
Values of the static friction coefficients for each of the materials used in this study are listed in table \ref{table1}.

\begin{table}[htdp]
\caption{Experimentally determined values of $\mu$ for indicated materials (abbreviations defined in text), and their elastic modulii (supplied by manufacturer).}
\begin{center}
\begin{tabular}{ccc}
Material & Friction Coefficient $\mu$ & Modulus E (GPa)\\
PE & 0.19 & 2.5\\
PEG & 0.14 & 2.5\\
PEA & 0.19 & 0.004\\
LEX & 0.22 & 2.0\\
TEF & 0.06 & 0.5\\
\end{tabular}
\end{center}
\label{table1}
\end{table}

\subsection{Experimental Protocol}
We initially place all disks in the interrogation chamber in random positions and move the boundary plates to an initial packing fraction chosen {\it a priori} to fall comfortably in an unjammed state.  The disks are subject to friction (with the glass bottom, the boundary and with each other) and the system jams below the random close packed density $\phi_{RCP} \approx 0.84$. After each quasi-static step, a 10 s time-trace of all six boundary sensors is collected at a sampling frequency of 1 KHz followed by a digital image of the whole system. The time-averaged value of the trace constitutes the pressure measured by the boundary sensor at a particular value of $\phi$. The boundaries are then moved through a quasi-static step ($\Delta \phi$), and the procedure is repeated. The only control parameter in this experiment is the quasi-static step $\Delta \phi$. No other external excitation or perturbation is applied. As a consequence, this experiment studies the pack evolution at ``zero granular temperature''. Unlike ideal jamming, real-world experiments have to contend with friction. Usually, frictional effects are circumvented by applying a granular temperature via acoustic excitation or vibration of boundaries that relax frictional stresses. In the absence of any such mechanism in this experiment, frictional effects become fully manifest. The ideal jamming requirements of zero temperature and zero friction are, experimentally, mutually incompatible. Meeting the zero friction requirement violates the zero temperature requirement and vice versa. Secondly, we do not tap the system after each quasi-static step to mimic annealing in simulations \cite{MSLB2007}. Tapping the system in experiments (or annealing in simulations via gradient minimization or alternative methods), no matter for how short a duration, or how weak in amplitude, is tantamount to application of an effective granular temperature which violates the zero temperature requirement. In order to understand the role of granular temperature, we performed one experimental run, discussed below, where the system is subjected to gentle (but systematically uncontrolled) tapping after each quasi-static step.

\section{Results}
\subsection{First Loading Cycle: Jamming Onset in presence of friction}

\begin{figure}
\begin{center}
\includegraphics[width = 3.3 in]{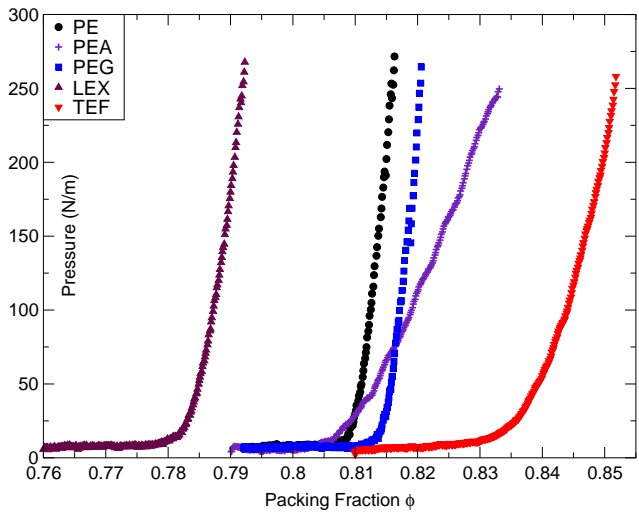}
\end{center}
\caption{(Color online) $P$ vs.  $\phi$ (quasi-static step-size $\Delta \phi = 1 \times 10^{-4}$). The packing fraction $\phi$ at which system pressure starts increasing monotonically falls with increasing static friction coefficient $\mu$. The change in pressure slopes arises from different elastic modulii of the materials used.}
\label{fig:fig2}
\end{figure}

Here we discuss the role of friction on the onset of jamming when the pack is compressed the first time from a random initial configuration. Silbert et al. have shown \cite{SEGHL2002, Silbert2010} that the critical coordination number $Z_c$ as well as the critical packing fraction $\phi_c$ at which jamming occurs shift to progressively lower values with increasing friction coefficient, owing to the additional structural stability provided by the tangential forces. Analysis for jamming \cite{SEGHL2002} and for the unjamming transition \cite{Silbert2010} reached similar conclusions. In fig. \ref{fig:fig2} we plot the pressure $P$ versus packing fraction $\phi$ for disks with different friction coefficients ($Z$ is not measured for reasons discussed in Section III A). Treating commencement of pressure rise as the jamming onset condition, we confirm the numerical prediction of Silbert et al. that jamming onset does occur at lower packing fractions with increasing friction. Whereas jamming onset also represents the jamming transition in {\it Ideal Jamming}, the same is not true in this experiment. Here the term ``jamming onset'' marks the nucleation of the first jammed cluster within the system. Unlike {\it Ideal Jamming} which is marked by an abrupt transition, in this experiment the jamming transition proceeds smoothly through a stress percolation mechanism, as we discuss in Section IV B.

The range of friction coefficients $0.06 \le \mu \le 0.22$ for disks employed in this study yields friction-dependent values of the critical packing fraction $\phi_c$ that are greater than the asymptotic steady value corresponding to random loose packing ($\phi_{RLP}$); one would need higher $\mu$ to reach RLP conditions.  Nevertheless, the experimental values of $\phi_c$ we measure for the experimental values of $\mu$ are close to those obtained in numerical simulations of Silbert et al. (see fig. 1 and table 1 of \cite{Silbert2010}).  Since we use different materials with varying elastic moduli, the slopes of the pressure curves vary between the materials. As a counterpoint, a quick comparison of PE and PEG data shows that the two plots have the same slope since they have the same modulus. Since PEA disks are softer, however, they have a shallower slope in comparison to PE data, but the pressure rise does approximately coincide for both PE and PEA data since they share the same friction coefficient. The slight difference in the $\phi$ value of pressure rise start between PE and PEA can be attributed to configuration dependent fluctuations that naturally arise between different experimental runs.

\begin{figure}
\begin{center}
\includegraphics[width = 3.1 in]{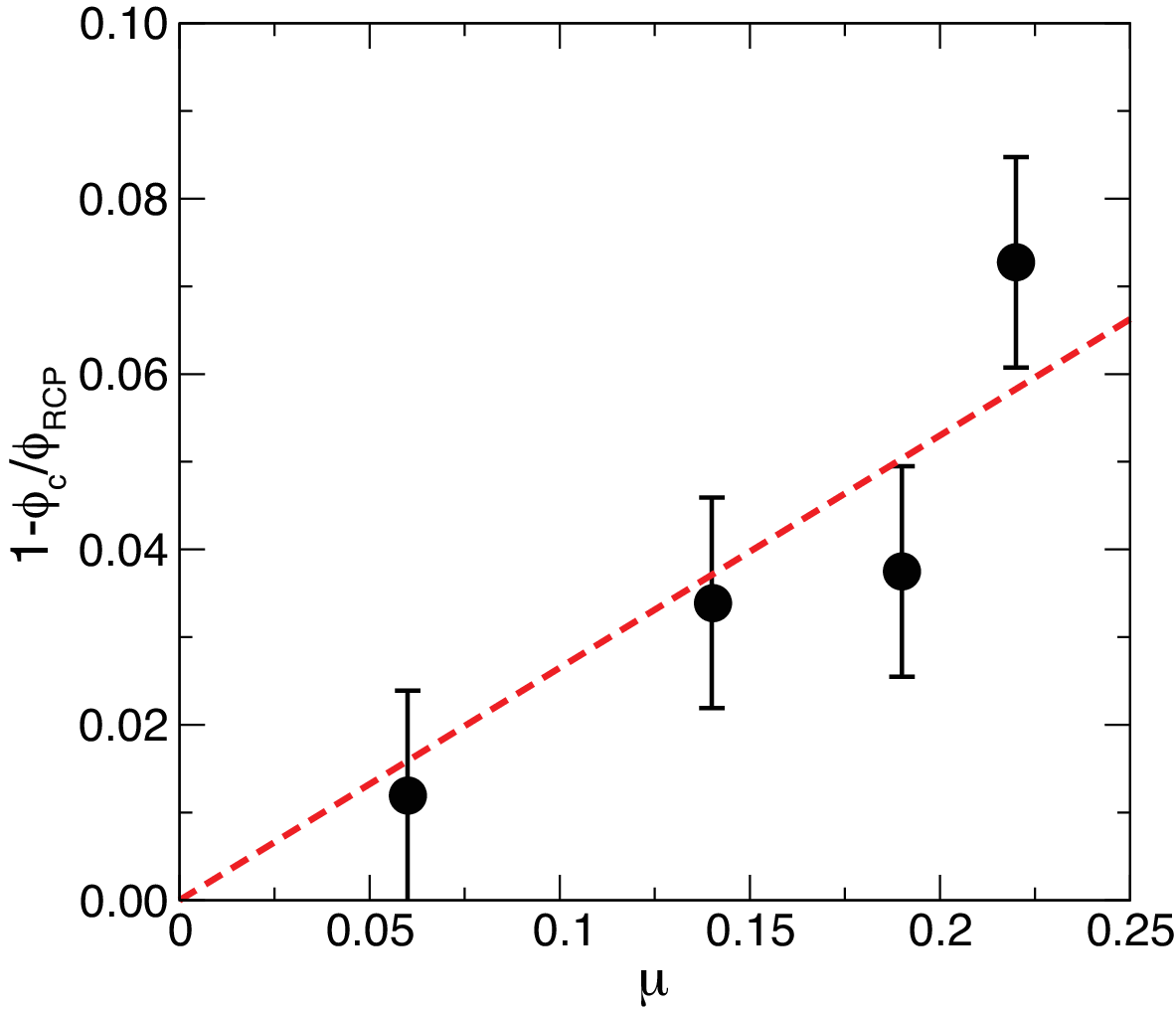}
\end{center}
\caption{(Color online) Deviation of fractional $\phi_c$ with respect to $\phi_{RCP}$ vs.  $\mu$ for materials described in fig.\ \ref{fig:fig2}. The error bars are estimates as opposed to statistical averages over many runs.}
\label{fig:fig3}
\end{figure}

In fig.\ \ref{fig:fig3}, we show the variation of the critical packing fraction $\phi_c$ as a function of the disk friction coefficient $\mu$.  In particular, the quantity $1-\phi_c/\phi_{RCP}$, the fractional deviation from the expected zero friction random close packed value $\phi_{RCP}$, increases consistent with a linear dependence on $\mu$, again agreeing with numerical simulations \cite{Silbert2010}.

\subsection{First Loading Cycle: Fragile Behavior}

\begin{figure}
\begin{center}
\includegraphics[width = 3.3 in]{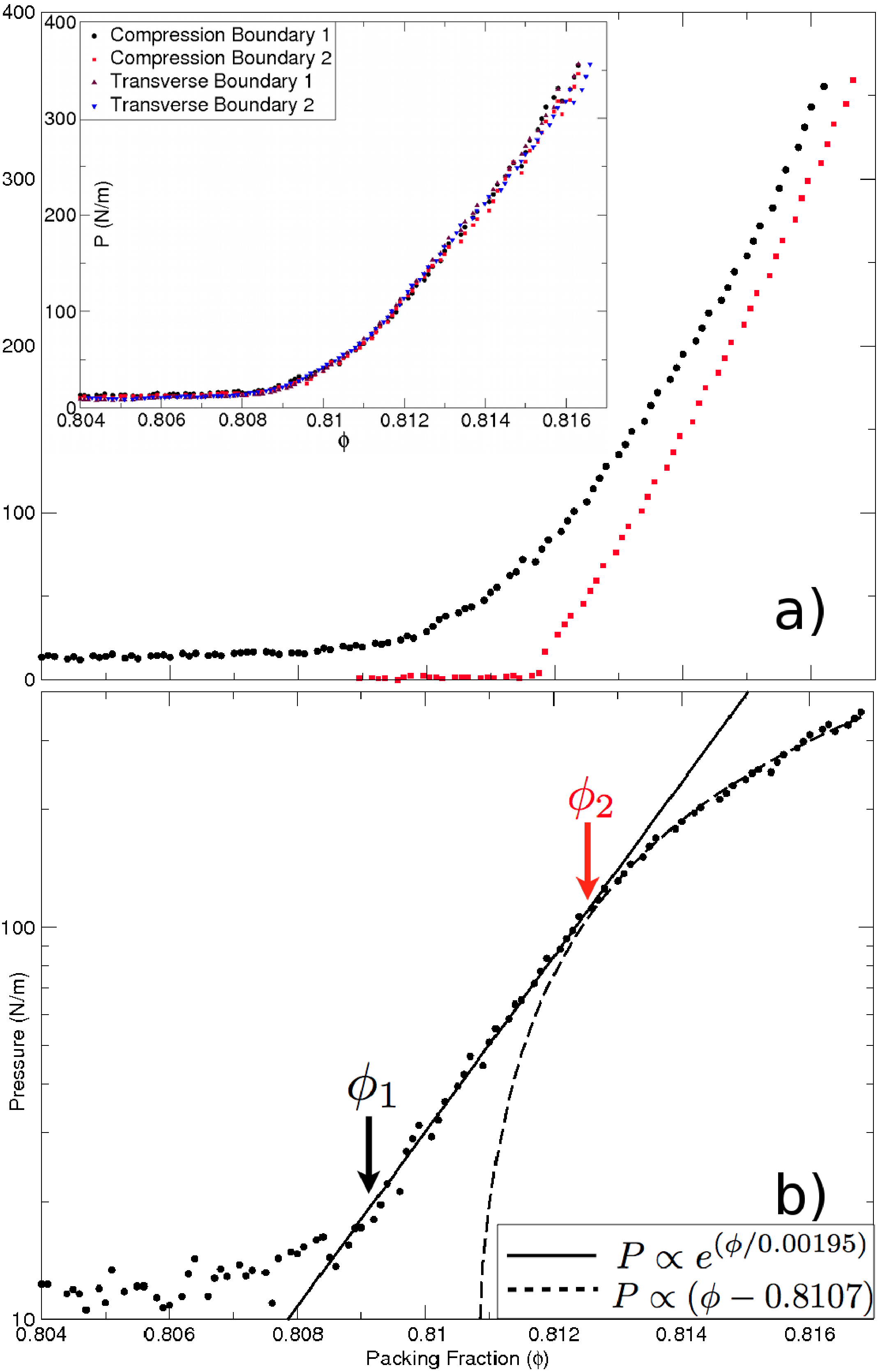}
\end{center}
\caption{(Color online) (a) $P$ vs. $\phi$ - linear scale - for the first (solid circles) and second (solid squares) compression cycles. The pressure scaling is gradual for the first cycle as compared to the more abrupt transition during the second cycle where $\phi_c$ is indicated. The inset plots the pressure from all four boundaries (same vertical and horizontal scale). No anisotropy is observed in the pressure signal. (b) $P$ vs. $\phi$ for the first jamming cycle - log-linear scale - shows the existence of an intermediate regime where pressure scales exponentially. The solid line is an exponential fit to the data $P \propto e^{\phi/\chi_P}$ with $\chi_P = 0.00195$, and the dashed line is a linear fit. The values of $\phi_1$ and $\phi_2$ are indicated in the plot.}
\label{fig4}
\end{figure}

Let us now consider the shape of the pressure curve itself by comparing the pressure signal for the first and second jamming cycles. Figure \ref{fig4}(a) plots $P$ from one compression boundary sensor versus $\phi$ for the first and second jamming cycles for Photoelastic disks (PE). The continuous increase in $P$ for the first cycle is qualitatively different from the more abrupt change in slope for the second cycle. The lateral shift in $P$ is a signature of a friction-induced hysteretic response which progressively shifts $\phi_c$ to higher values as the system is repeatedly jammed and unjammed (to be discussed below). The vertical shift in $P$ within the flat (unjammed) regime for the first cycle (black solid circles in fig. \ref{fig4}a) is traced to the friction between the movable boundary plates plus the mobilized fraction of disks and the glass bottom. At the end of the first loading cycle, when the system is decompressed, the boundary plates only move back until stresses in the pack are relieved. Further quasi-static reverse stepping of micrometers does not cause continued backward motion of boundary plates to their initial positions. Instead, the micrometers decouple (loss of physical contact) from the boundary plates. For this reason, the unjammed regime observed during the second compression cycle (red solid squares in fig. \ref{fig4}a) does not show the vertical shift in $P$. During the first cycle, the boundaries move in and push the disks towards a jammed configuration. When unjammed and jammed again, the contacts that were developed at the end of the first jamming cycle are re-activated, and the stresses build up as the system is subjected again to uniaxial compression.

In order to better understand the smooth increase in $P$ for the first jamming cycle, fig.\ \ref{fig4}(b) plots $P$ versus $\phi$ on a log-linear scale. One sees three distinct regimes in the pressure curve. In the unjammed (consolidation regime) $P$ is essentially flat, modulo a shallow ramp due to friction between disks and bottom glass plate. The second regime is characterized by an exponential increase in $P$ beginning at $\phi_1 \approx 0.8093$. The solid line in fig.\ \ref{fig4}(b) is an exponential fit to the data $P \propto e^{\phi/\chi_P}$ with $\chi_P=0.00195$. Note that although the range of $\phi$ for the exponential regime is small, the increase in $P$ over that interval is almost one decade. The pressure eventually deviates from this exponential regime and settles to a regime that seems to fit well with linear scaling (dashed line in fig. \ref{fig4}(b)) $P \propto \phi$ at $\phi_2 = 0.8124$. Owing to limited range of pressure data for $\phi > \phi_2$, it is not clear whether this regime scales linearly or algebraically ($P \propto \phi^{\psi}$) with the exponent $\psi$ being marginally greater than 1. The inset in fig.\ \ref{fig4} (a) plots $P$ from all four boundaries against $\phi$ on a linear scale. The remarkably good collapse of the pressure curves atop one another strongly suggests that anisotropic effects arising from uniaxial compression are not detected by the pressure sensors. This isotropy may result from our geometry for which the aspect ratio is $L/W = 0.54$ and would probably not persist for large aspect ratios, i.e., $L/W >> 1$.

Although this exponential pressure scaling for the first loading cycle seems to contradict the predicted \cite{OSLN2003, HC2005, MSLB2007} power-law for $P$ across the jamming transition, as we will discuss below, it can be explained through a stress percolation mechanism. The predicted power-law scaling (approximately linear \footnote{The best fit of the data to the form $P \propto (\phi-\phi_c)^{\beta}$ yields $\beta = 1.1$, consistent with \cite{MSLB2007}}) is, however, recovered for subsequent jamming cycles as shown for the second jamming cycle in fig.\ \ref{fig4}a where $\phi_c = 0.8118$ is determined by the point at which $P$ starts to rise.

\begin{figure}
\begin{center}
\includegraphics[width = 3.3 in]{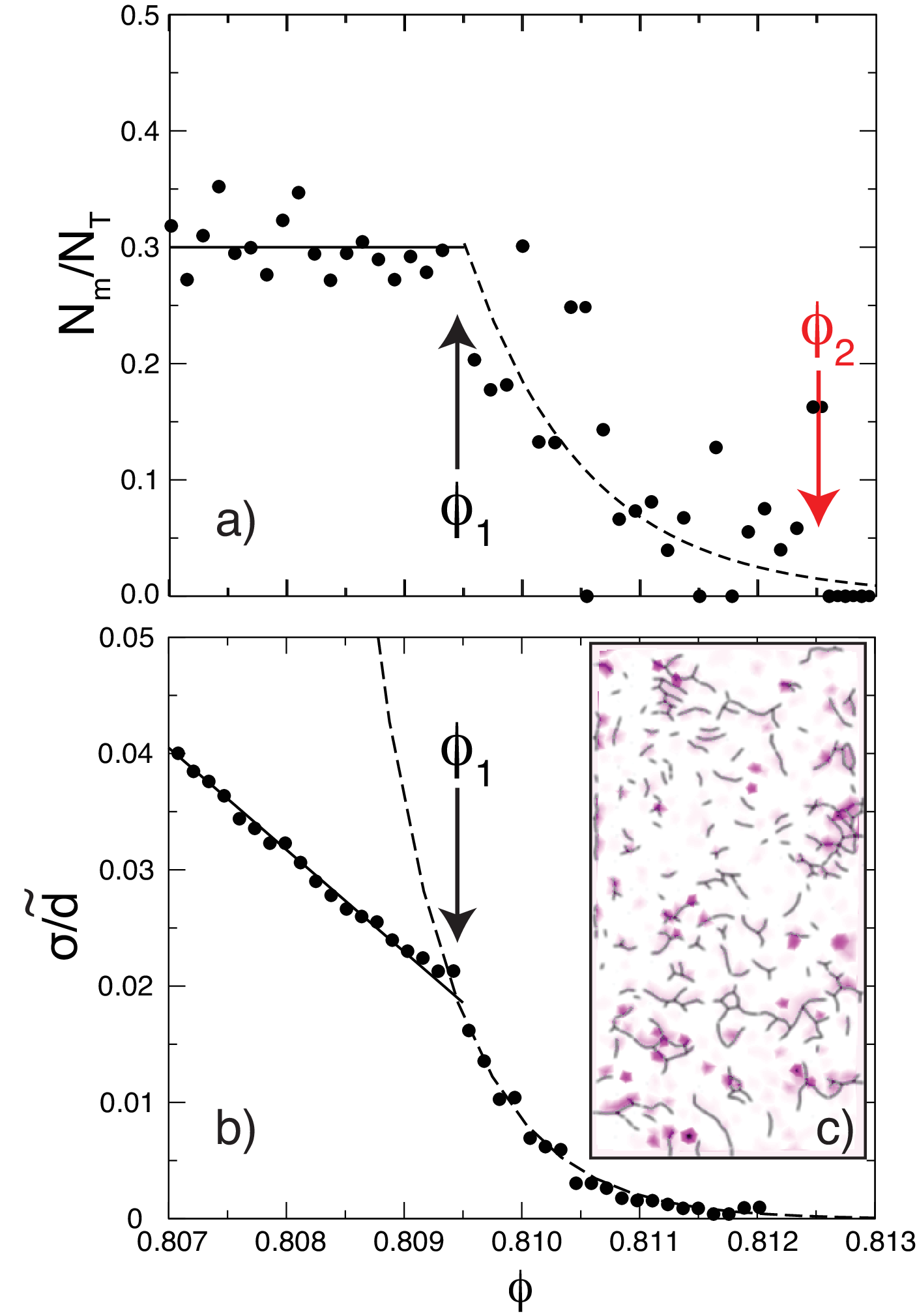}
\end{center}
\caption{(Color online) (a) Fraction of disks moving $N_m/N_T$ vs $\phi$ showing constant region for $\phi < \phi_1$ and exponential decay for the fragile regime ($\phi_1<\phi<\phi_2$) with $\chi_N \approx 0.001$. (b) Normalized displacement variance $\sigma/\tilde{d}$ relative to a state near $\phi_2$ showing a linear decrease for $\phi < \phi_1$ and an exponential decrease with $\chi_{\sigma} = 0.0007$ for the fragile regime ($\phi_1<\phi<\phi_2$). (c) Image of the superposition of the difference in the stress network (black lines) and the magnitude of disk displacements between $\phi = 0.8095$ and $\phi = 0.8105$. Arrows indicate $\phi_1$ and $\phi_2$.}
\label{fig5}
\end{figure}

We now look at the behavior of disk displacements.  Figure \ref{fig5}(a) shows the fraction of disks moving a distance greater than about 1\% of a mean disk diameter \footnote{There are smaller displacements below our threshold that may behave differently as a function of $\phi$.} as a function of $\phi$ where $N_m$ is the number moving and $N_T=1900$ is the total number of disks.  The fraction is constant at about $0.3$ up to $\phi_1 = 0.8094$, after which it decreases rapidly up to the jamming value $\phi_2 = 0.8124$.  The decrease is consistent with an exponential $e^{-\phi/\chi_N}$ with $\chi_N \approx 0.001$.  The variance of individual disk displacements $\sigma$ (normalized by $\tilde{d} = (d_L+d_S)/2 = 0.794$ cm) relative to a state near the jamming threshold \footnote{Picking different reference states near or beyond jamming causes a finite offset in variance of different magnitudes, but the exponential decay is preserved with the same value of $\chi = 0.0007$ to within experimental precision.} is shown in fig.\ \ref{fig5}(b) as a function of $\phi$. One again observes two distinct regimes: a linear one that corresponds to the unjammed (consolidation) regime where the pressure curve in fig.\ \ref{fig4}(b) is flat, and a second one in which the variance decreases exponentially with $\sigma \propto e^{-\phi/\chi_\sigma}$ with $\chi_{\sigma} = 0.0007$. This exponential drop in displacement variance occurs over the same interval in $\phi$ as the exponential regime of the pressure curve in fig.\ \ref{fig4}(b) as indicated by the arrows depicting $\phi_1$ and $\phi_2$.  The presence of disk displacements in the exponential regime, where a percolating force network already exists, suggests that the particles that are part of the force network still undergo small displacements and deformations, which in turn allows visible displacement {\it inside} the network region, eventually leading to the refining of the network.  This picture is reinforced by the data in the inset fig.\ \ref{fig5}(c) where the difference in the force chain network between $\phi = 0.8095$ and $\phi = 0.8105$ is shown as black lines and the spatial distribution of the magnitude of particle displacements is also shown. The correlation of new stress chain creation (the differences) and the particle displacements is striking.

\begin{figure}
\begin{center}
\includegraphics[width = 3.3 in]{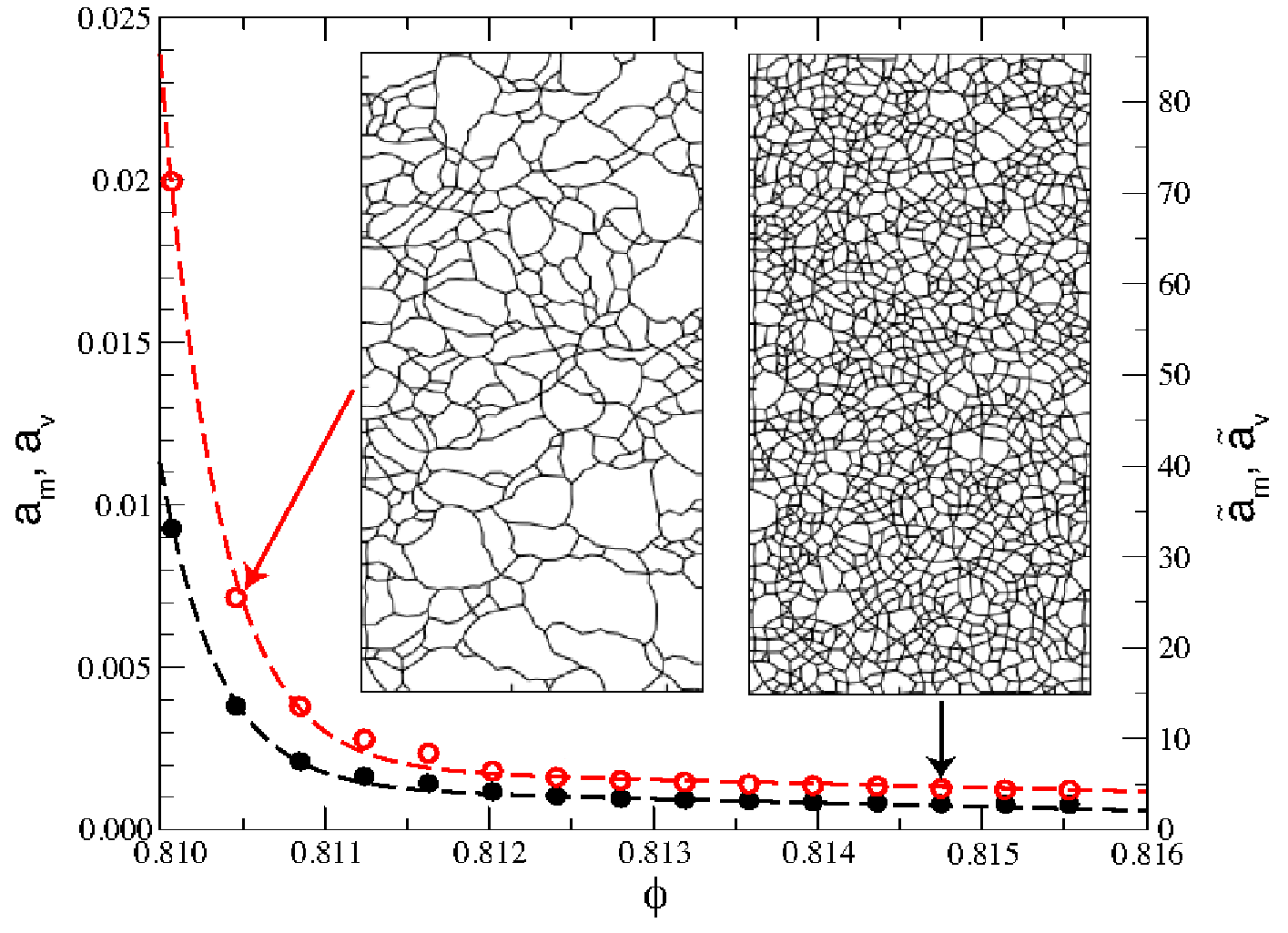}
\end{center}
\caption{(Color Online) Mean ($\bullet$) and variance ($\circ$) of stress chain domain area normalized by total area, $a_{m}$, $a_{v}$ (left axis) or by the mean area formed by connections between different combinations of 3 disks in a close packed triangular array, $\tilde{a}_{m}$, $\tilde{a}_{v}$ (right axis), respectively. Insets show stress networks and corresponding domains for $\phi = 0.8104$ (left) and $\phi= 0.8148$ (right). The dashed lines are fits to a combined linear dependence with a decaying exponential (see main text).}
\label{fig6}
\end{figure}

To further study the nature of the different regimes observed in global pressure measurements, we consider local properties of disk configurations, namely the structure of stress chains and the measurements of disk displacements.  The identified stress chains enclose domains with no measured stress as illustrated in the insets of fig.\ \ref{fig6}.  We characterize the stress chain networks by the mean $a_m$ and variance $a_v$ of the fractional domain area (relative to the total cell area on the left axis) and also by the mean $\tilde{a}_m$ and variance $\tilde{a}_v$ of the domain area but now normalized (right axis) by the average of the area enclosed by a triangular arrangement of small and large disks (all large, all small, two large, two small).  As indicated in the inset images, the mean and variance decrease rapidly between 0.810 (the lowest value of $\phi$ for which we could determine the stress chain network) and $\phi_2 = 0.8124$.  For higher $\phi > 0.8124$, the mean
and variance decrease linearly with a small slope.  Fits to an exponentially decreasing function $e^{-\phi/\chi_s}$ as shown in fig.\ \ref{fig5} yields the same value of $\chi_s = 0.00035$.  The exponential decrease in $a_m$ and $a_v$ is another signature of the fragile jammed state.

\subsection{Repetitive Loading: Friction-induced Hysteresis and Creep}

\begin{figure}
\begin{center}
\includegraphics[width = 3.3 in]{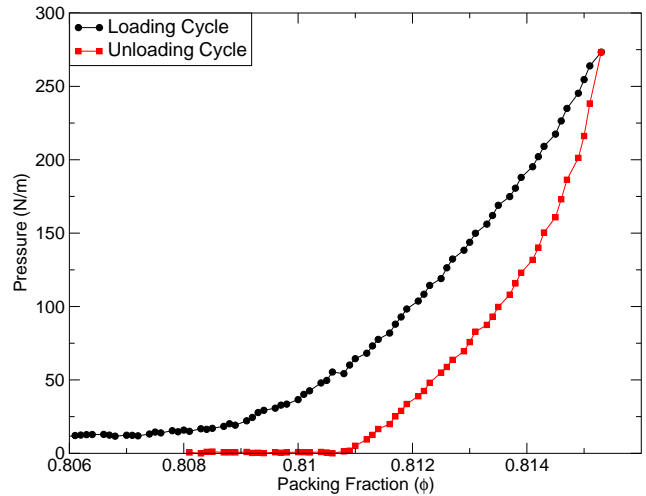}
\end{center}
\caption{(Color Online) $P$ vs. $\phi$ (equivalent to stress-strain measurement) for PE disks at $\Delta \phi = 1 \times 10^{-4}$ exhibits hysteresis.}
\label{fig7}
\end{figure}

Following the first loading cycle, we now turn our attention to the role of friction under repetitive loading and unloading. Here we report two frictional effects that strongly deviate from current jamming predictions. When the granular pack is subjected to loading and unloading, hysteresis is observed in the pressure versus packing fraction plot (see fig.\ \ref{fig7}). If subjected to repetitive loading-unloading cycles, the system exhibits creep whereby the packing fraction at which the system jams progressively shifts and the hysteresis curves evolve towards higher packing fractions. Friction-induced hysteresis \cite{Zhang2010}, as well as the rate-dependent behavior of $\phi_{c}$ \cite{IOS2011} have been reported recently. Because frictional jamming is heavily dependent upon preparation protocol, we are unable to offer a comparison between our study and prior works. Nevertheless, since our experimental setup design shares close correspondence with standard mechanical load cell designs, it allows us to compare our results with relevant amorphous solids in the geophysical context (e.g., certain forms of sandstones and sedimentary rocks).

\begin{figure}
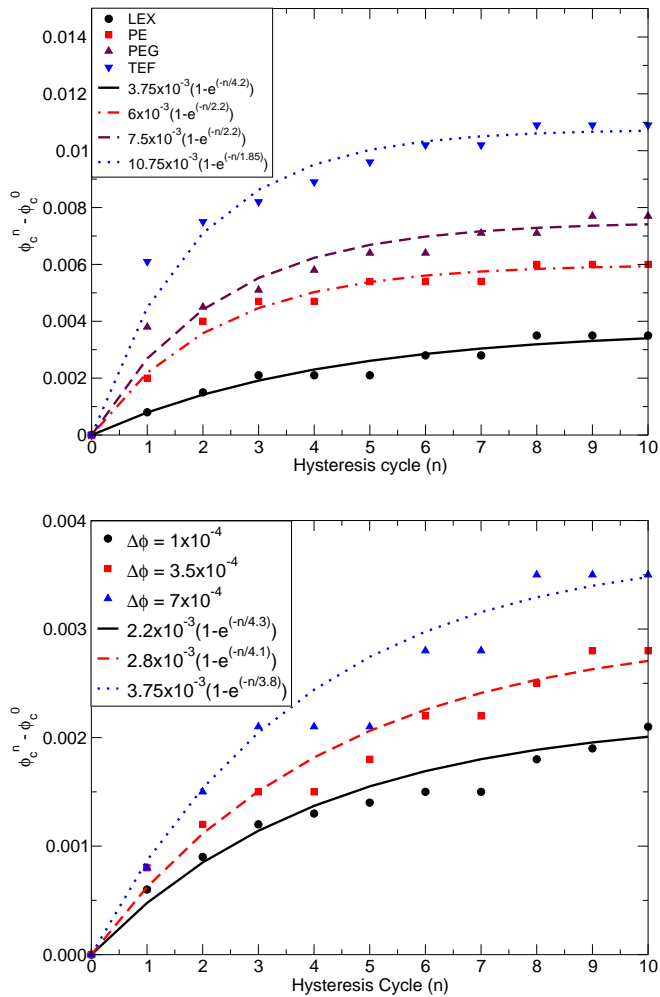

\begin{center}$
\begin{array}{cc}
\includegraphics[width = 3.4 in]{fig8a.eps}\\\\
\includegraphics[width = 3.4 in]{fig8b.eps}
\end{array}$
\end{center}
\caption{(Color online) Top: Difference between $\phi_c^n$ and its initial onset value $\phi_c^0$ as a function of the number of cycles of loading and unloading $n$ for different materials. Bottom: Same quantities for the PE material and for different step sizes $\Delta \phi$. Materials and step sizes
are labeled in the plot legends as are the coefficients in the solid-line exponential fits.}
\label{fig8}
\end{figure}


In fig. \ref{fig8} we plot the difference in $\phi_c$ between the $n^{th}$ ($\phi_c^n$) and $0^{th}$ ($\phi_c^0$) loading-unloading cycles against the cycle number $n$. In the top panel we vary the friction coefficient $\mu$ while keeping the quasi-static step-size constant at $\Delta \phi = 1 \times 10^{-4}$, whereas in the bottom panel we keep the friction coefficient constant at $\mu = 0.19$ for PE disks, while varying the quasi-static step-size $\Delta \phi$. The rate at which the pack evolves from $\phi_{c}$ to $\phi_{RCP}$ exhibits monotonic dependence on the friction coefficient with the pack evolution being quickest for TEF with lowest friction coefficient, followed by PEG, PE, and finally LEX with the highest friction coefficient. Also as shown in fig. \ref{fig8} bottom panel, the quasi-static step-size $\Delta \phi$ which controls the magnitude of perturbation provided to the pack also controls the pack evolution rate monotonically at step-sizes $\Delta \phi = 1 \times 10^{-4}, 3.5 \times 10^{-4},$ and $ 7 \times 10^{-4}$, thereby exhibiting rate dependence in the pack evolution in a quasi-static sense.

Given the uncertainty in the effect of parasitic friction, it is difficult to be confident about the functional form of the relaxation.  Significant care would be required to tease out these relationships accurately but the qualitative dependence on $\mu$ and $\Delta \phi$ seems solid.

\section{Discussion}
Having presented the results, we now discuss them vis-a-vis ideal granular jamming,  explain how they are related to each other, and establish how they are related to prior works. We start with the first compression cycle where fragile behavior is observed. The critical packing fraction $\phi_c$ in ideal jamming is that packing fraction at which two conditions are simultaneously met: (i) the packing fraction at which pressure starts rising above zero, and (ii) the isostatic point where total degrees of freedom equal the total number of constraints, i.e., the number of floppy modes equals $D(D+1)/2$ (rigid body rotation and translation), and all individual disk displacements are strongly impeded. The first condition is met at $\phi_1$, the packing fraction at which pressure starts rising above zero (albeit exponentially and not power-law). The second condition is met, however, only at $\phi_2$ where all displacements become very small; ergo the two conditions defining $\phi_c$ are well separated by a fragile, exponential regime characterized by simultaneous existence of non-zero pressure (jammed clusters) and non-zero displacements (unjammed clusters). As noted in Section II A, however, the definition of a contact in ideal jamming permits the system exist in any one of two discrete states - completely unjammed, or completely jammed; an intermediate regime as demonstrated by the fragile state is not allowed. This anomalous behavior therefore raises several questions vis-a-vis the ideal jamming paradigm, which we explain below:

1) Why have prior studies that have successfully verified the jamming predictions not reported this anomalous scaling behavior? All prior experimental \cite{MSLB2007, Zhang2005} and numerical \cite{OSLN2003} studies to our knowledge study the unjamming transition, i.e., they approach $\phi_c$ from the jammed state towards the unjammed state. An analysis of the unjamming over jamming transition is favored for technical reasons. Precise detection of pressure rise commencement around $\phi_c$ is very difficult to detect over numerical/instrumental noise and fluctuations from discrete configurational adjustments during compression. We therefore believe the fragile state exists in their systems, but forms part of the experimental preparation phase which may not have been systematically analyzed.

Two exceptions lend support to this possibility. In recent numerical work on contact percolation transition (CPT), Shen et al. \cite{SOS2012} analyze the approach to jamming transition and show deviations (discussed below) occur prior to jamming onset. But perhaps of greater relevance to the present study is the earlier experimental work of X. Cheng \cite{C2010} where the jamming transition was studied by swelling tapioca pearls in water. Of particular interest is fig. 14 of \cite{C2010} where the structural factor (pair correlation), measured boundary force, and mean square particle displacements are plotted against packing fraction. The force exhibits two distinct regimes at packing fractions labeled $\phi_1$ and $\phi_2$, where $\phi_2$ is shown to coincide with random close packing. The pair correlation function exhibits two distinct peaks at $\phi_1$ and $\phi_2$. Finally, the mean square displacement goes through a maximum between $\phi_1$ and $\phi_2$ and falls to zero at $\phi_2$. This behavior is related to existence of local jammed clusters starting at $\phi_1$ which grow until global jamming is achieved at $\phi_2$. That study traces the source of this anomalous behavior to friction, the proof in support being vibrational disturbances (non-zero granular temperature) relieve these frictional contacts and recover ideal jamming predictions (see fig. 16 in \cite{C2010} and related discussion). The role of friction and vibrational disturbance in the present study is presented later in this article. Given frictional jamming exhibits sensitive dependence on experimental protocol and preparation history, the correspondence between Cheng's experiment and the present study is noteworthy, particularly since they follow different experimental protocols.

2) Does an alternative physical mechanism explain the fragile state? In the fragile regime, part of the system is jammed as evidenced by non-zero pressure (fig. \ref{fig4}b), while the remainder is unjammed as evidenced by non-zero displacements (fig. \ref{fig5}). The co-incidence of exponential pressure rise and fall in displacements with increasing $\phi$ points to the percolation of jammed clusters across the system. Evidence of stress percolation comes in two parts: firstly, the strong spatial correlation between local disk displacements and nucleation of new stressed contacts (fig. \ref{fig5}c), and secondly, exponential decrease in the fractional area enclosed by stress chains (fig. \ref{fig6}). The present work is not isolated in its claim of a percolation route to jamming. Several recent studies \cite{SOS2012, CBD2012, BZCB2011} have shown some of percolation mechanism preceding the jamming transition.

\begin{figure}
\begin{center}
\includegraphics[width = 3.3 in]{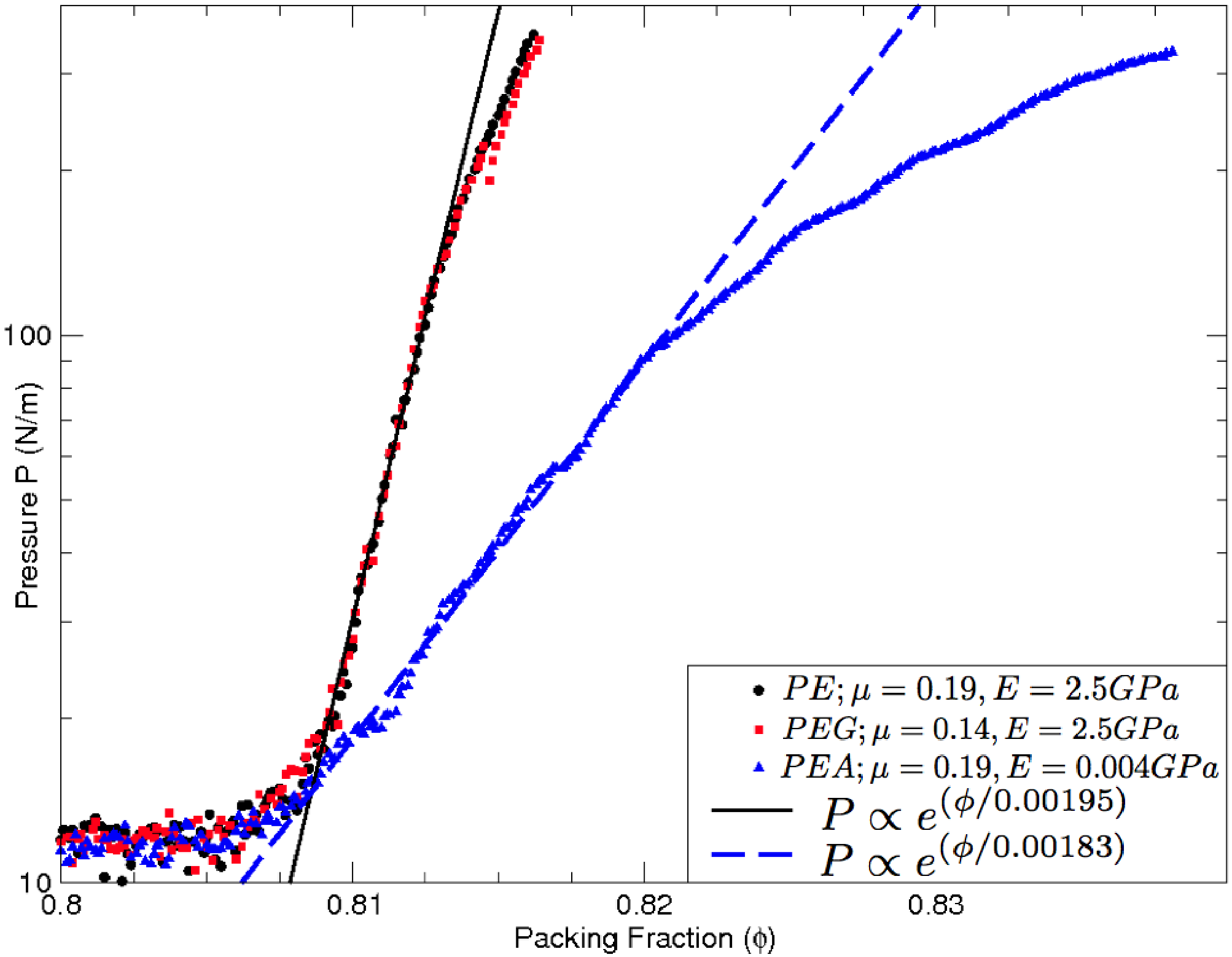}
\end{center}
\caption{(Color Online) $P$ vs. $\phi$ in log-linear scale for the first compression cycle. All plots have been horizontally shifted for coincidence of $\phi_1$. The exponential pressure scaling ($P \propto e^{\phi/\chi_P}$) for PE and PEG exhibits same slope ($\chi_P = 0.00195$) as demonstrated by the exponential fit (solid black line), implying variation in friction coefficient has no effect on $\chi_P$. On the other hand, PEA disks with same friction coefficient as PE, but lower modulus exhibit a shallower exponential scaling (with $\chi_P = 0.0063$, long dashed blue line) over approximately same range of pressure as PE and PEG.}
\label{fig9}
\end{figure}

3) How does friction control the fragile state? Figure \ref{fig:fig2} presents clear evidence that the start of fragile state (or $\phi_1$) is directly dependent upon the friction coefficient. With increasing value of the friction coefficient, the fragile state commences at a lower packing fraction. In fig. \ref{fig8} we replot the data presented in fig. \ref{fig:fig2} in log-linear scale for PE, PEG and PEA disks. The data for PEG and PEA are horizontally shifted so their $\phi_1$ values coincide with that of PE disks. It is apparent from fig. \ref{fig8} that PE and PEG disks have the same exponential slope ($P \propto e^{\phi/\chi_P}$, $\chi_P = 0.00195$). They share the same modulus ($E = 2.5$ GPa) but different friction coefficients ($\mu = 0.19$ for PE, $\mu = 0.14$ for PEG) suggesting friction has no measurable effect on the slope of exponential pressure scaling.

In contrast, the fragile regime for PEA disks which have same friction coefficient as PE disks ($\mu = 0.19$) but have much lower modulus ($E = 0.004$ GPa) show shallower scaling with $\chi_P = 0.0063$ implying $\chi_P$ depends upon the modulus. This does not mean, however, friction has no effect on $\chi_P$. There are in fact two modulii entering the pressure measurement, modulus of the disk material and the effective pack modulus. We have only varied the disk material modulus, the pack modulus on the other hand is a function of coordination number, which in turn depends upon the friction coefficient. Hence, the effective pack modulus is nonlinear due to coordination number; a fact also evident from power-law scaling of $P$ versus $\phi$ curves, equivalent to Stress-Strain relations, in Ideal Jamming. From that relation ($P \propto (\phi-\phi_c)^{\psi}$) one can discern the nonlinear effective modulus must be $\psi-1$). We note two subtleties that arise here. Firstly, all experimental, and most numerical $P$ versus $\phi$ curves are measured for finite systems, and the asymptotic approach of $\psi$ in the thermodynamic limit (large system size) is not understood. Albeit subtle, nonlinear elastic constants play a central role in elasto-plastic responses of amorphous solids \cite{KLP2010, HKLP2011}. Furthermore, since the Coulomb yield criterion ($F_T \le \mu F_N$) only provides a lower bound on the value of the tangential stress component, it is not possible to experimentally or theoretically learn how friction controls the coordination number, and therefore the effective pack modulus. Empirical deduction from numerical simulations may be able to shed some light on this relationship. In light of this, one can only say that friction controls the exponential slope $\chi_P$ indirectly via effective pack modulus, but cannot explain how.

4) Why is fragile behavior not observed during subsequent compression cycles? When we decompress the system after the first compression cycle, the stresses in the pack are relieved and the boundaries move just enough to relax the system. The disks, however, are left in the final configuration in which they found themselves at the end of the first compression cycle. If the granular pack is subjected to a second compression, the contacts that existed at end of the first compression cycle are immediately activated everywhere across the system simultaneously at a critical packing fraction $\phi_c$. This situation exactly corresponds to the sudden system wide emergence of stressed contacts at $\phi_c$.

\begin{figure}
\begin{center}$
\begin{array}{cc}
\includegraphics[width = 3.2 in]{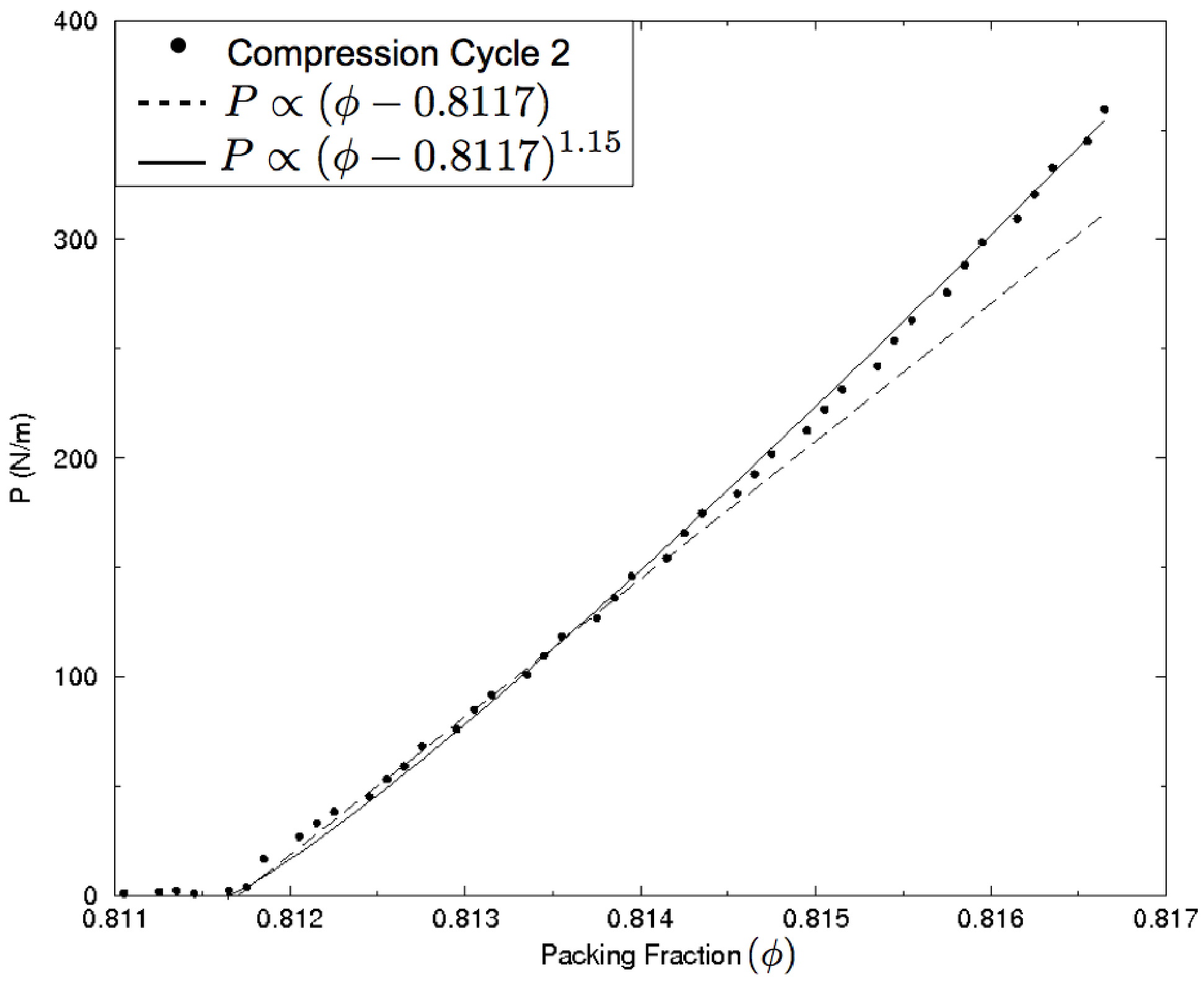}\\
\includegraphics[width = 3.2 in]{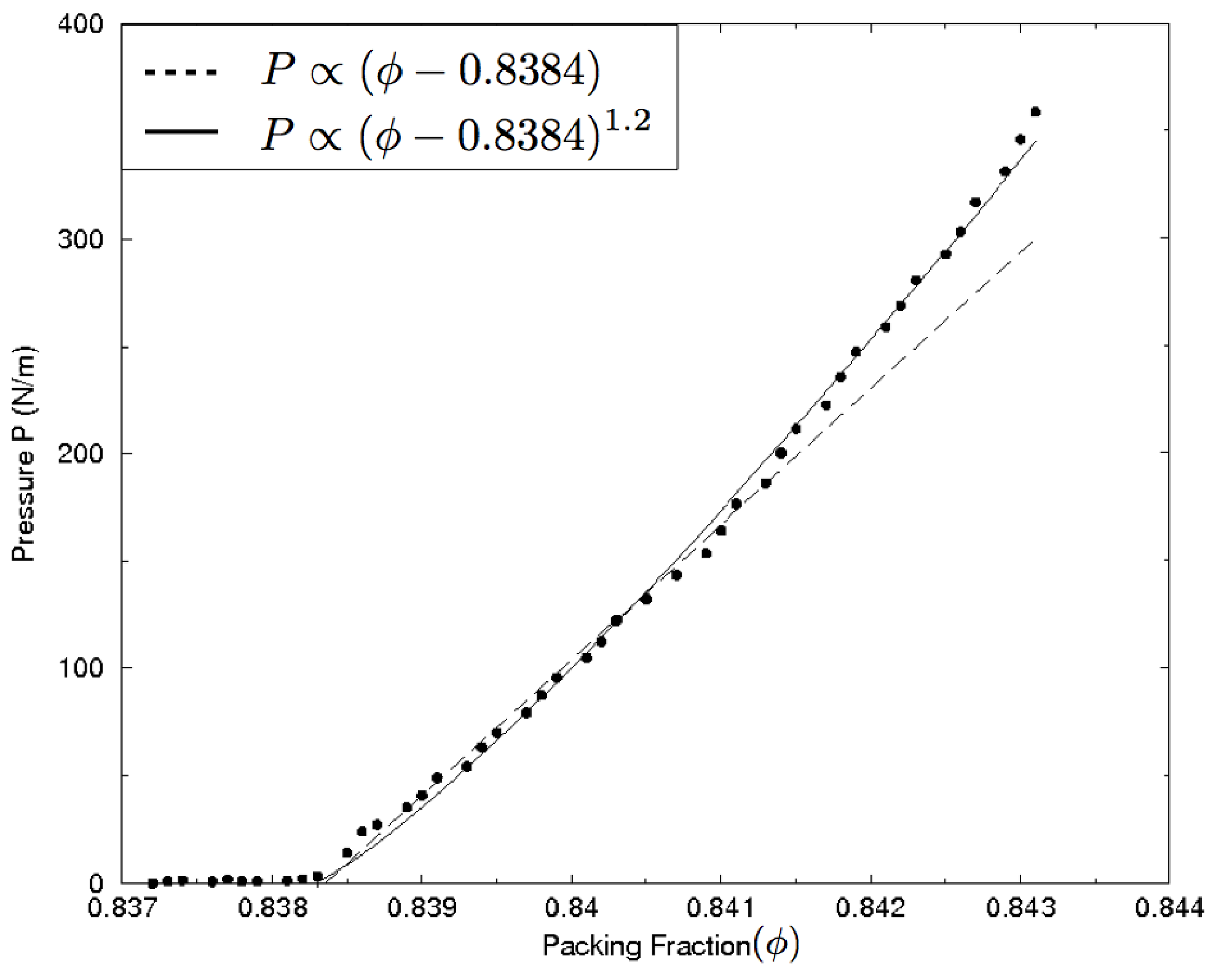}
\end{array}$
\end{center}
\caption{Top: $P$ vs. $\phi$ for the second compression cycle in linear scale shows the predicted power law scaling for granular jamming transition. The pressure rise is abrupt, and is in contrast with the gradual (exponential) rise observed for the first compression cycle. The solid line is the best power law fit to the experimental data with a power law exponent of 1.15. The long-dashed line representing the best linear fit to the data demonstrates that the experimental data does not follow linear scaling. Bottom: $P$ vs. $\phi$ for first compression cycle where the system is gently tapped (albeit with no systematic control) after each quasi-static step.}
\label{fig10}
\end{figure}

In fig. \ref{fig10} top panel, we plot the pressure $P$ against the packing fraction $\phi$ for the second compression cycle. We recall that Jamming theory predicts zero pressure below $\phi_c$. At $\phi_c$ when the system satisfies the isostatic condition, and all constraints are activated simultaneously across the system, a rise in pressure is recorded with a power-law scaling ($P \propto (\phi-\phi_c)^{\psi}$. Prior experiments by Majmudar et al. \cite{MSLB2007} have demonstrated the power-law increase in pressure with an exponent of 1.1. Our experimental data are fit very well with an exponent of 1.15 (see solid line fit for the experimental data in fig. \ref{fig7}), and are in very good agreement with the results in \cite{MSLB2007}. Given that Majmudar et al. tapped their system after each quasi-static step, which we do not, the agreement in the exponent is indeed remarkable.

As a control, we also conducted one experimental run where the system was subjected to gentle (but systematically uncontrolled) tapping after each quasi-static step during the first compression cycle for multiple reasons. Firstly, we wanted to verify that our granular system reproduces the results of Majmudar et al. \cite{MSLB2007} since the essential point of departure between their measurements lies in the protocol - namely, tapping the system to mimic annealing. Accordingly, we were able to recover their result for the jamming transition. We do note that the $\phi_c$ is evidently still below $\phi_{RCP} = 0.84$ as predicted for 2D systems. We attribute the discrepancy to two possibilities. Firstly, we do not systematically control the tapping process. We merely tapped the system along the four boundaries gently with a mallet after each quasi-static cycle. Hence the plot in the bottom panel of fig. \ref{fig10} only serves as qualitative verification. Secondly, the discrepancy can also be attributed to our system size, $N = 1900$ disks.  Following from the numerical studies of O'Hern et al. \cite{OSLN2003}, where $\phi_c$ was shown to be a configuration dependent random variable for finite system size, it is conceivable that the control experiment we performed found itself in a configuration that jammed marginally below $\phi_{RCP}$. Finally, as noted in Section II A, the location of $\phi_c$ relative to $\phi_{RCP}$ not withstanding, the scaling properties about $\phi_c$ are robust.

The control run for the first loading cycle with tapping is an excellent example of how the presence of friction makes pack evolution strongly protocol dependent where in the absence of external perturbations the pack exhibits exponential scaling, but with tapping it exhibits power-law scaling. In addition, this control run also points to another subtle feature of the jamming paradigm. The ideal jamming predictions are predicated on the requirements of zero friction, zero temperature, and zero applied stress. The results observed with and without tapping show us zero friction and zero temperature are incompatible requirements in the real-world where one cannot escape frictional effects save in strong exceptions like jamming in foam. One then begs to ask, if tapping \cite{MSLB2007} is considered equivalent to annealing processes invoked in simulations, is the zero temperature requirement being adhered to? Also, if one interprets tapping as a thermal kick (rather than a constant thermal agitation), it is tantamount to destroying the system's evolution history after each quasi-static step. Then are the ideal jamming predictions a result of a protocol that renders the pack memoryless? We believe these are important issues that merit further work, since thermalization can act both to render a system memoryless as in the present instance, or help to retain system memory \cite{KN2011}.

The incompatibility between zero friction and zero temperature raises another important question about what it means for a frictional granular pack to be structurally stable. As argued in \cite{SEGHL2002}, with increasing friction a granular pack can jam at $\phi_c < \phi_{RCP}$, and the isostatic point can occur at $D+1 \le Z_c \le 2D$. At zero temperature but non-zero friction, repetitive loading data exhibits evolution in $\phi_c$ which implies there must be an increase in $Z_c$ at each cycle which we are unable to measure owing to experimental shortcomings. Nevertheless, a straightforward physical interpretation for hysteretic creep may be presented from the granular jamming perspective. Owing to friction, the system jams at $\phi_c$ into a metastable configuration, and will remain so indefinitely unless perturbed externally (recall, the disks are macroscopic and not susceptible to thermal fluctuations). Any external driving (e.g., tapping or jiggling) relieves frictional stresses in the system (at least partially if not all of them) and destroys this metastable configuration. In the absence of such external drive, the only perturbative mechanism available to the system is the magnitude of the quasi-static step ($\Delta \phi$) which is equivalent to strain. Hence, it follows that with each loading-unloading cycle, the system evolves ever so slightly through a series of metastable configurations towards a final, stable configuration. The magnitude of $\Delta \phi$ therefore directly controls the pack evolution as shown in the bottom panel of fig. \ref{fig8}. Furthermore, whether or not a given value of $\Delta \phi$ can evolve the system from one metastable configuration to the next depends upon the static friction coefficient $\mu$ which controls the degree of a configuration's stability - the higher the friction coefficient, the more stable a configuration is. The rate at which $\phi_c$ evolves must therefore be a function of $\mu$ and $\Delta \phi$. Indeed, as shown in fig. \ref{fig8}, the difference ($\phi_c^n - \phi_c^0$) between $\phi_c$ for the $n^{th}$ and $0^{th}$ cycles depends upon $\Delta \phi$ and $\mu$.

Interestingly, strain dependent creep and hysteresis are also observed in viscoelastic materials which represent a totally different class of amorphous media. Unlike purely elastic substances, a viscoelastic substance has an elastic and a viscous component. Purely elastic materials do not dissipate energy when a load is applied and removed, but a viscoelastic medium does \cite{MeyersChawla}. Hysteresis is observed in the stress-strain curve, with the area of the loop being equal to the energy lost during the loading cycle. Since viscosity is the resistance to thermally activated plastic deformation, a viscous material will lose energy through a loading cycle. Plastic deformation results in lost energy, which is uncharacteristic of a purely elastic material's reaction to a loading cycle. More specifically, viscoelasticity is a molecular rearrangement. When a stress is applied to a viscoelastic material such as a polymer, parts of the long polymer chain change position. This movement or rearrangement is called Creep \cite{Ferry1980}. Polymers remain a solid material even as these parts of their chains are rearranging in order to accompany the stress, and as this occurs, it creates a back stress in the material. When the back stress is the same magnitude as the applied stress, the material no longer creeps. Unlike viscoelastic media where creep and healing of metastable polymer configurations are thermally activated, in the athermal granular system considered here, the quasi-static strain is the only perturbative mechanism available by which the system creeps towards its ultimate stable configuration at $\phi_{RCP}$. Whereas the dissipative mechanism available to viscoelastic amorphous media is supplied by viscosity, in the granular system it comes about through friction. Such viscoelastic behavior has been observed in naturally occurring granular packs in the geophysical context, namely sandstone and sedimentary rocks \cite{CKT2009, Tutuncu1998}. Particularly noteworthy is the fact that our loading protocol is very similar to loading procedures followed in measuring mechanical properties of geophysical rock samples in standard load cells.

\section{Summary}

In summary, we have presented experimental results for a system of bi-dispersed, frictional disks subjected to uni-axial compression. We verify the numerical predictions for frictional jamming \cite{SEGHL2002, Silbert2010} whereby jamming is shown to occur at progressively lower packing fractions with increasing friction coefficient. We also show the first compression cycle exhibits exponential increase in pressure, and a corresponding exponential fall in displacements over a range of packing fractions $\phi_1 < \phi < \phi_2$. We show this exponential scaling separates the two conditions that define the critical packing fraction $\phi_c$. We compare our data against published experimental and numerical results and delve into how friction controls this regime in a non-trivial manner. To put our work in perspective,  it falls within a class of recent results that demonstrate some form of percolation mechanism arising prior to jamming transition, with stress percolation presenting the route to jamming in the present case.  Finally, we find hysteretic creep under repetitive loading-unloading cycles and experimentally trace its source to friction. Despite our inability to reliably measure coordination numbers, our experiments help explain the various regimes arising in frictional granular jamming.

\acknowledgments 
This work was carried out under the auspices of the National Nuclear Security Administration of the U.S. Department of Energy at Los Alamos National Laboratory under Contract No. DE-AC52-06NA25396. The authors gratefully acknowledge helpful discussions with O. Dauchot.

\bibliography{all}

\end{document}